\documentclass[traditabstract]{aa}
\usepackage{color}

\def\obj{SDSS~J0013+1523}
\def \obja{SDSS~J0827+5224} 
\def \objb{SDSS~J0919+2720} 
\def \objc{SDSS~J1005+4016} 
\usepackage{graphicx}
\usepackage{setspace}
\usepackage{txfonts}
\usepackage{pstricks}
\include{epsf}
\usepackage{natbib}
\bibpunct{(}{)}{;}{a}{}{,}

\def\0{\phantom0}

\def\kms{km s$^{-1}$}
\def\kmsmpc{km s$^{-1}$ Mpc$^{-1}$}

\def\OIII{[\ion{O}{III}]}
\def\OII{[\ion{O}{II}]}

\begin{document}

\title{Three  QSOs acting as strong gravitational lenses{\thanks{
Based on data obtained at the W.M. Keck Observatory, which is  
operated as a scientific partnership among the California Institute of  
Technology, the University of California and the National Aeronautics  
and Space Administration. The Observatory was made possible by the  
generous financial support of the W.M. Keck Foundation. Also based on observations 
made with the NASA/ESA Hubble Space Telescope, obtained at the Space Telescope 
Science Institute, which is operated by the Association of Universities for Research in Astronomy, 
Inc., under NASA contract NAS 5-26555. These observations are associated with program \#GO12233. 
}}}


\titlerunning{Strong gravitational lensing by QSOs}

\author{The authors}

\author{F.~Courbin\inst{1} \and C.~Faure\inst{1} \and S.G.~Djorgovski\inst{2, 3} \and F.~R\'erat\inst{1} \and M.~Tewes\inst{1} 
\and G.~Meylan\inst{1} \and D.~Stern\inst{4} \and \\ A.~Mahabal\inst{2}\and 
T. Boroson\inst{5}, R.~Dheeraj\inst{6} \and D.~Sluse\inst{7, 8}}

   \institute{Laboratoire d'astrophysique, Ecole Polytechnique
     F\'ed\'erale de Lausanne (EPFL), Observatoire de Sauverny,
     CH-1290 Versoix, Switzerland
     \and
     Division of Physics, Mathematics, and Astronomy, California Institute of Technology, 
     Pasadena, CA 91125, USA
     \and
     Distinguished Visiting Professor, King Abdulaziz University, Jeddah 21589, Saudi Arabia
     \and     
     Jet Propulsion Laboratory, California Institute of Technology, Mail Stop 169-221, 
     Pasadena, CA 91109, USA
     \and
     National Optical Astronomy Observatory, Tucson, AZ 85719, USA
     \and
      The University of Maryland, College Park, MD 20742, USA
      \and
    Astronomisches Rechen-Institut am Zentrum f\"ur Astronomie der Universit\"at Heidelberg, 
    M\"onchhofstrasse 12-14, 69120 Heidelberg, Germany
    \and
    Argelander-Institut f\"ur Astronomie, Bonn Universit\"at, Auf dem H\"ugel 71, D-53121 Bonn, Germany
  } 

   \date{Received; accepted }
 
\abstract{We report the discovery of three new cases of QSOs acting as strong gravitational lenses on background emission 
line galaxies: 
\obja\ ($z_{\rm QSO}=0.293$, $z_{\rm s}=0.412$), \objb\ ($z_{\rm QSO}=0.209$, $z_{\rm s}=0.558$), \objc\ ($z_{\rm QSO}=0.230$, $z_{\rm s}=0.441$). 
The selection was carried out using a sample of 22,298 SDSS spectra displaying at least four emission lines at a redshift beyond that of 
the foreground QSO. The lensing nature is confirmed from Keck imaging and spectroscopy, as well as from 
HST/WFC3 imaging in the F475W and F814W filters. Two of the QSOs 
have face-on spiral host galaxies and the third is a QSO+galaxy pair. The velocity dispersion of the host galaxies, inferred from simple lens modeling,
is between $\sigma=210$ and $285$ \kms, making these host galaxies comparable in mass with the SLACS sample of early-type strong lenses. 

\keywords{Gravitational lensing: strong  --
          Galaxies: quasar: individual (\obj, \obja, \objb, \objc)   }}
\maketitle
%

\section{Introduction}
\label{intro}

Strong gravitational lensing is now a standard tool to study the mass distribution in galaxies. 
About 250 cases of galaxies acting as strong lenses on a background
object (quasar or galaxy) are known to date and have been observed with  HST 
\citep{Munoz1998, Bolton2008, Faure2008}. This growing sample of lenses allows us to 
measure the total radial mass profile of galaxies using the combined constraints given by the
shape of the lensed source and the dynamics of the lensing galaxy 
\citep[e.g.,][]{Auger2010, Barnabe2009} and to probe sub-structures in lensing
galaxies 
\citep[e.g.,][]{vegetti2010, chantry2010, yoo2005, biggs2004, kochanekdalal2004, keeton2003, koopmans2002, mao1998}. When the lensed source is photometrically
variable, like a QSO, the time delay constraint can also be used \citep{Refsdal1964} to measure
the Hubble parameter, $H_0$, independently of any standard candle \citep[e.g.,][]{Suyu2010}. 

Recently, we started building a new sample of lenses where the lensing object is a QSO host
galaxy. Our goal is to weigh QSO host galaxies for the first time with strong gravitational lensing
and to compare their dynamical and lensing masses. We presented the search method in \citet{Courbin2010a} 
which included the first case of a QSO acting as a strong lens, based on Keck Adaptive Optics 
(AO) imaging observations. In the present paper, we show three new strong `QSO lenses" for which we have Keck optical 
imaging and spectroscopy as well as deep Hubble Space Telescope (HST) images in two  
bands with the WFC3 camera in the visible channel. Our sample, when completed, will consist in a tool to test directly the 
scaling laws established between the properties of quasars emission lines, the mass of the central black hole, and the total mass of the host galaxies \citep[e.g.,][]{KAS, BON, SHEN}.

Throughout this paper,  the WMAP5 $\Lambda$CDM cosmology  is assumed ($\Omega_{\rm m}=0.258$, 
$\Omega_\Lambda=0.742$, $H_0=72$ \kmsmpc). All magnitudes are in the AB system.

\begin{figure*}[p!]
\begin{center}
\includegraphics[width=16.cm, height=7.5cm]{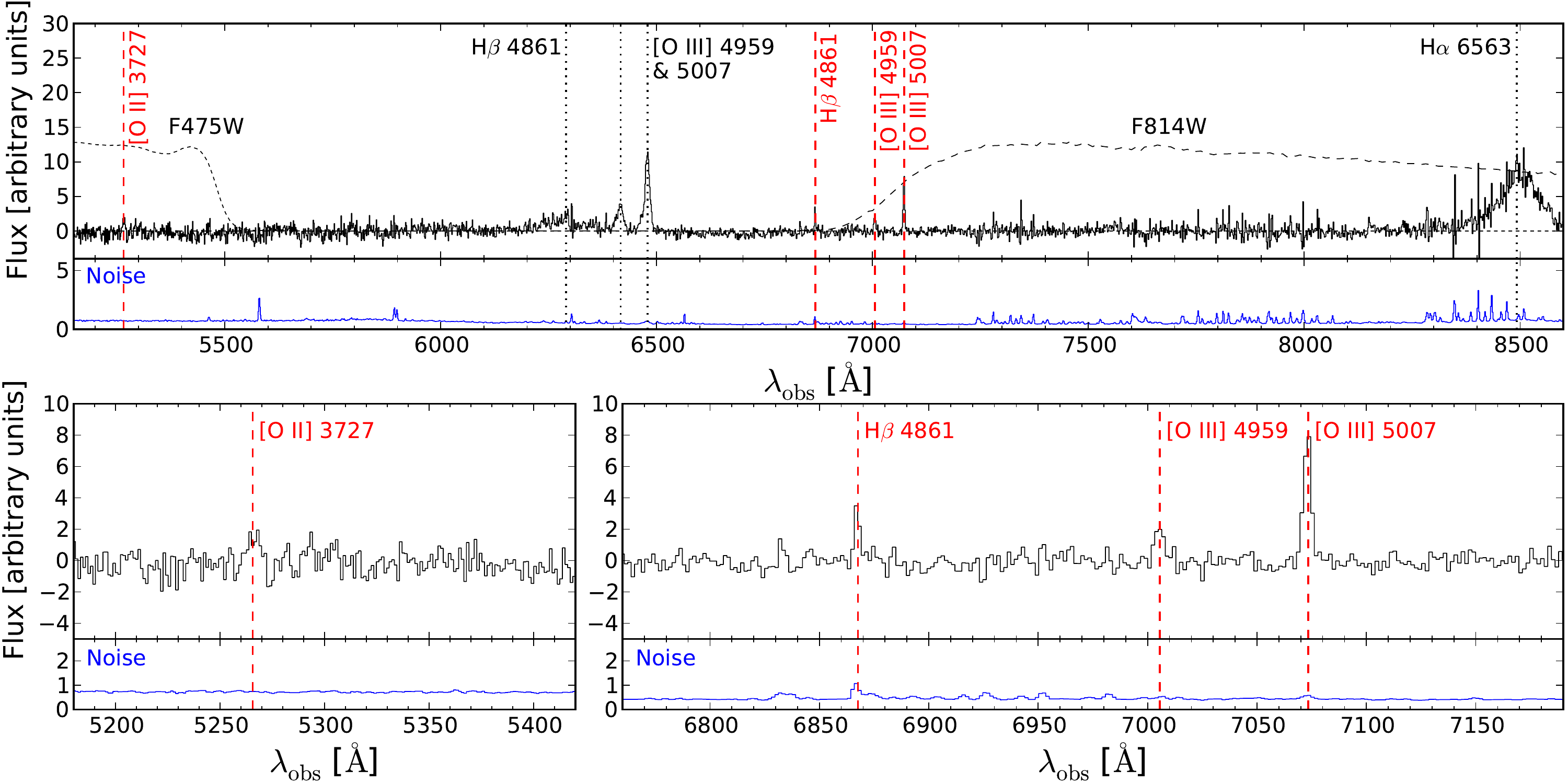}
\vskip 15pt
\includegraphics[width=16.cm, height=7.5cm]{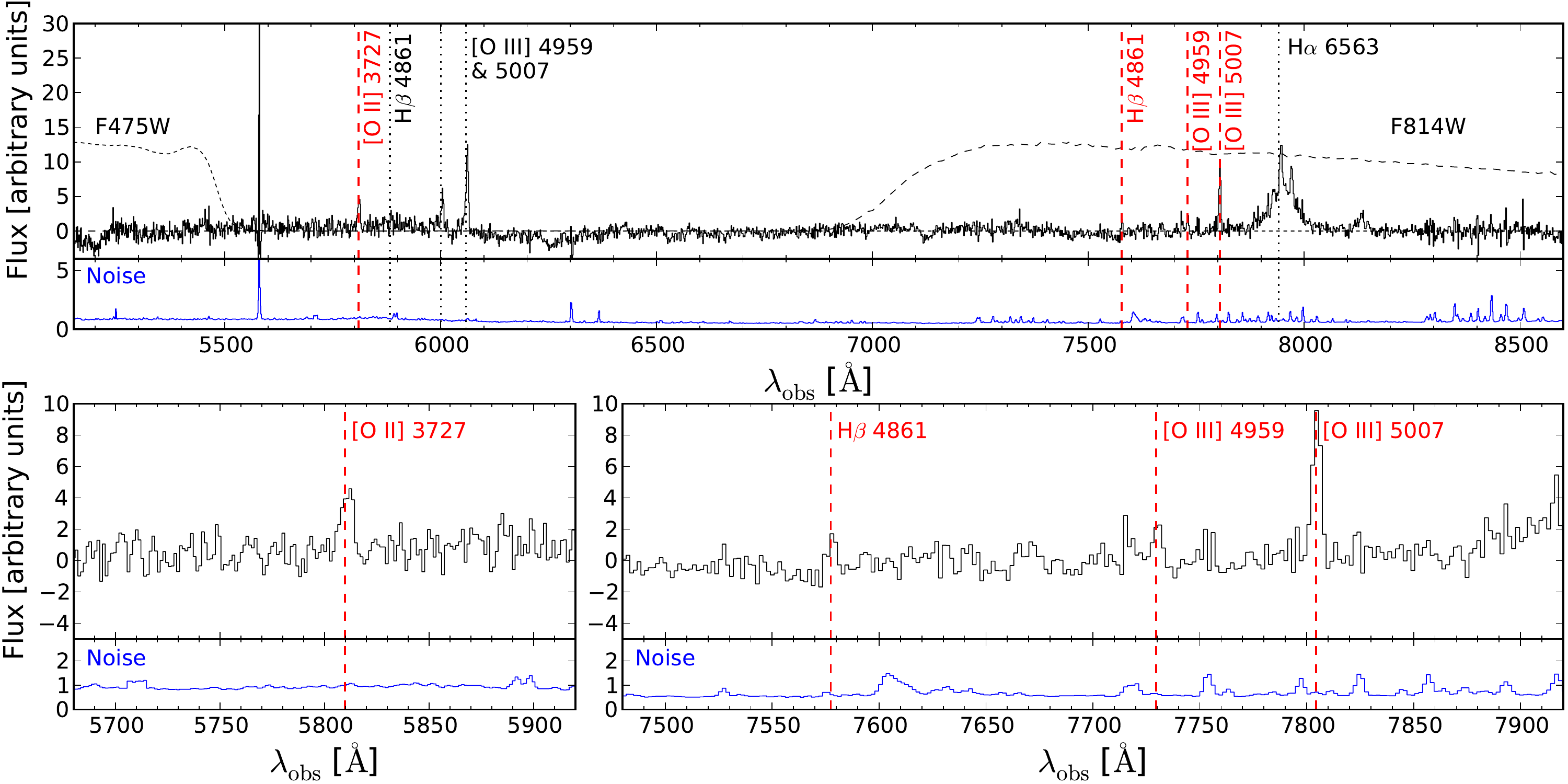}
\vskip 15pt
\includegraphics[width=16.cm, height=7.5cm]{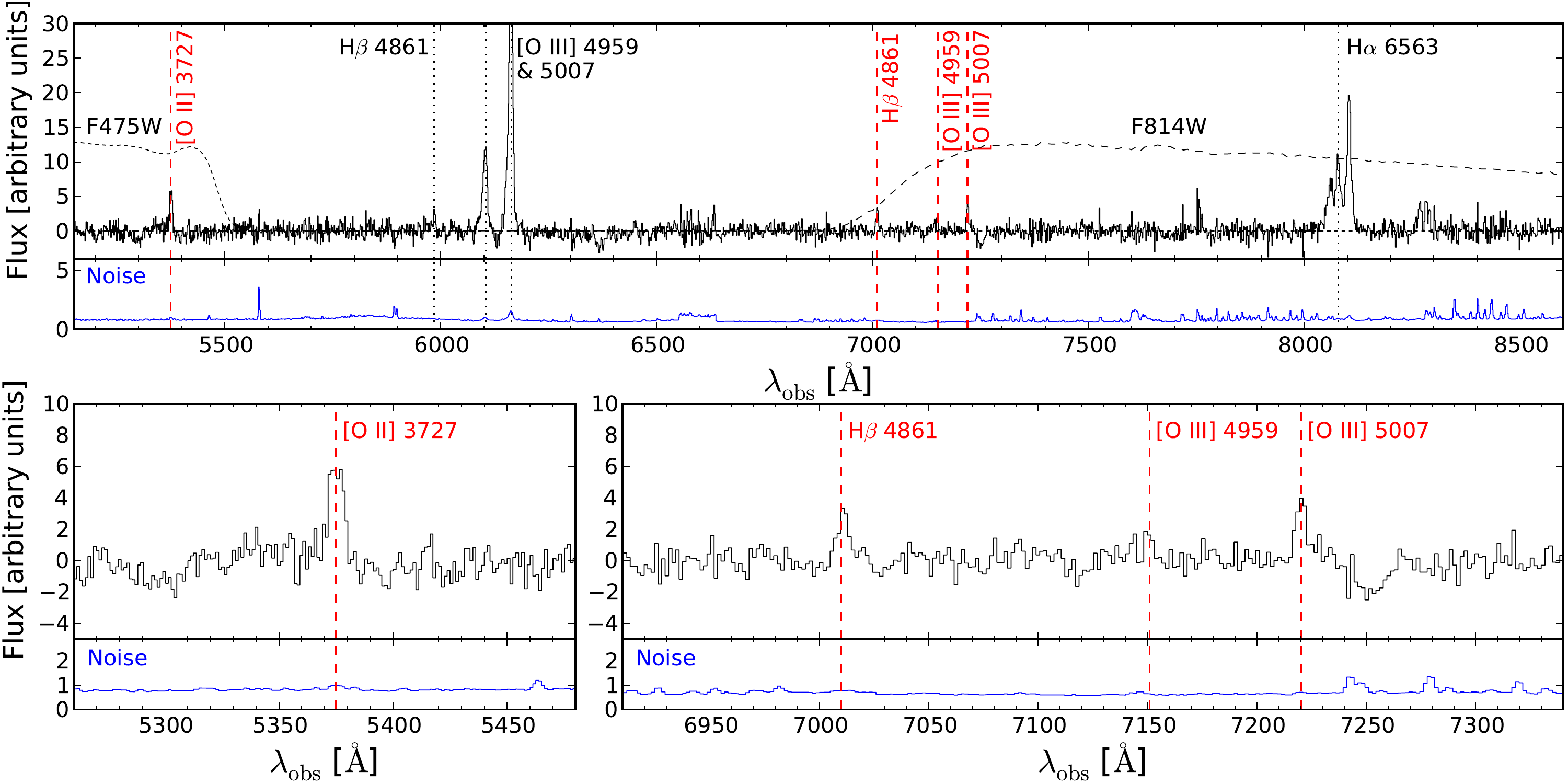}
\caption{SDSS spectra of the three new QSO lenses. From top to bottom are 
 \obja, \objb, and \objc. The emission lines of the foreground QSOs are indicated with
black vertical lines (dotted) and the emission lines of the background objects are shown with red vertical lines (dashed). 
All three QSOs have four significant background emission lines. All spectra are shown in the observed frame. The HST filter
curves are also displayed.}
\label{fig:SDSS_spectra}
\end{center}
\end{figure*}

\begin{figure*}[ph!]
\begin{center}
\includegraphics[width=16.cm, height=8cm]{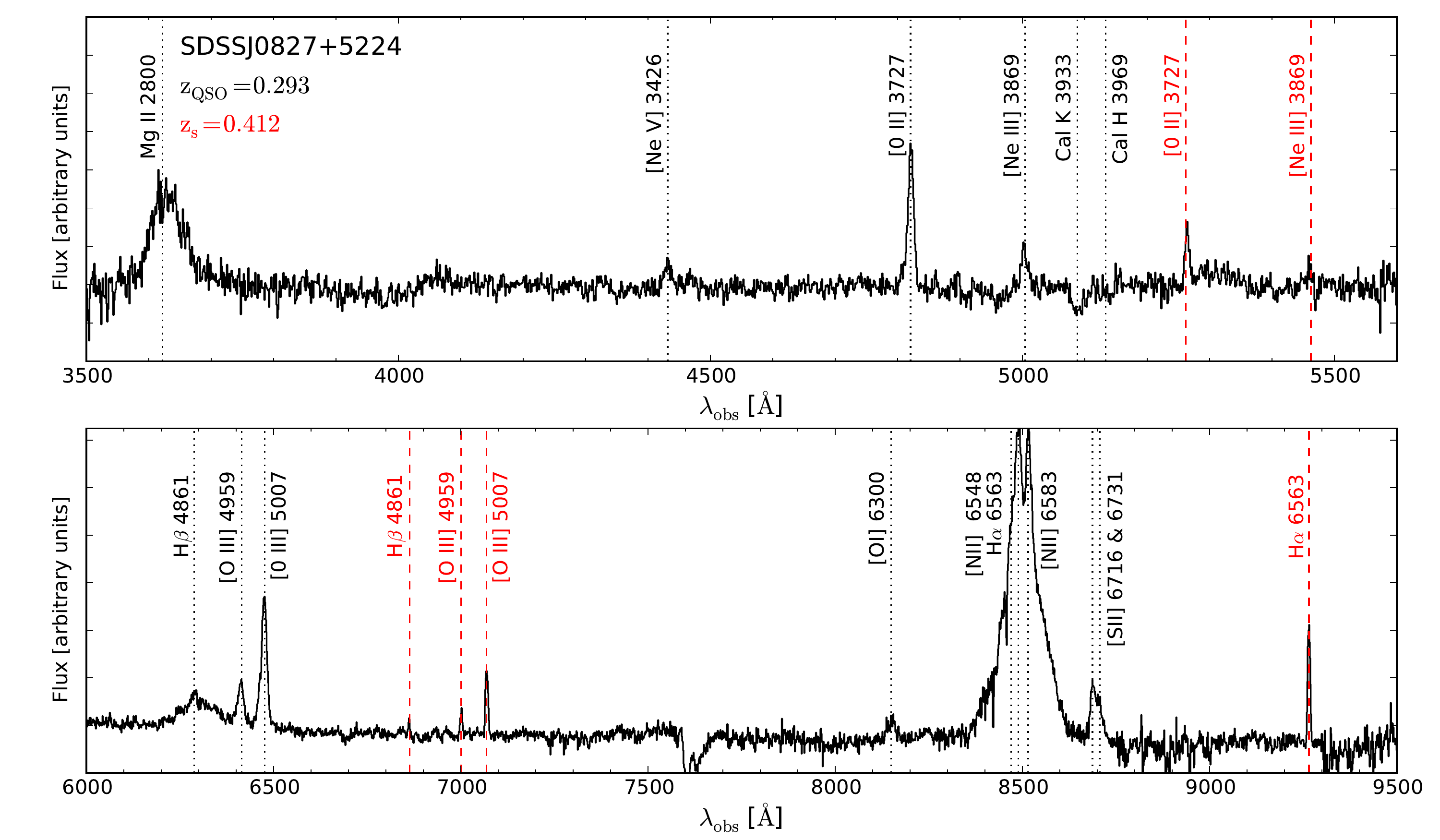}
\vskip 10pt
\includegraphics[width=16.cm, height=8cm]{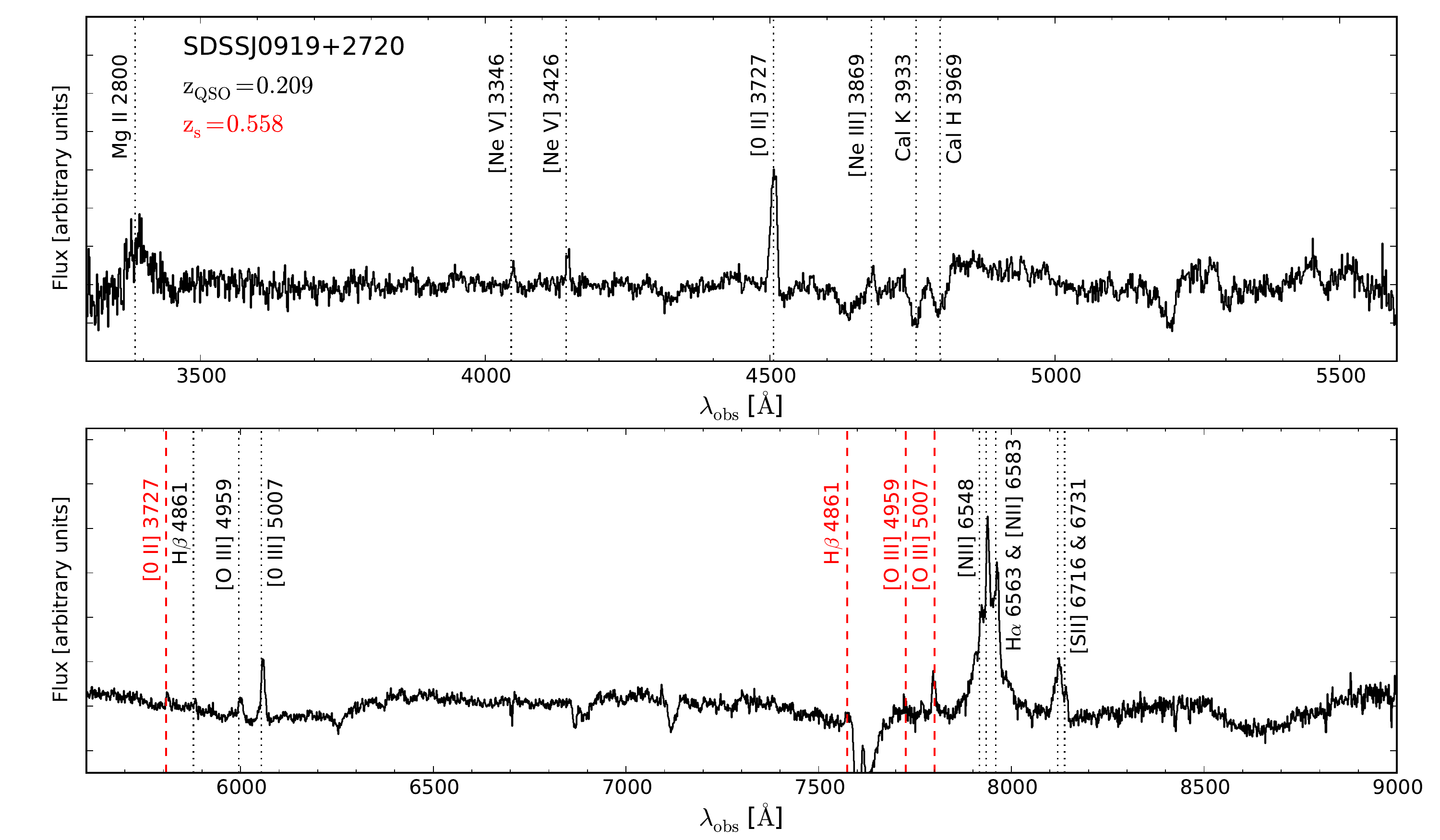}
\vskip 10pt
\includegraphics[width=16.cm, height=8cm]{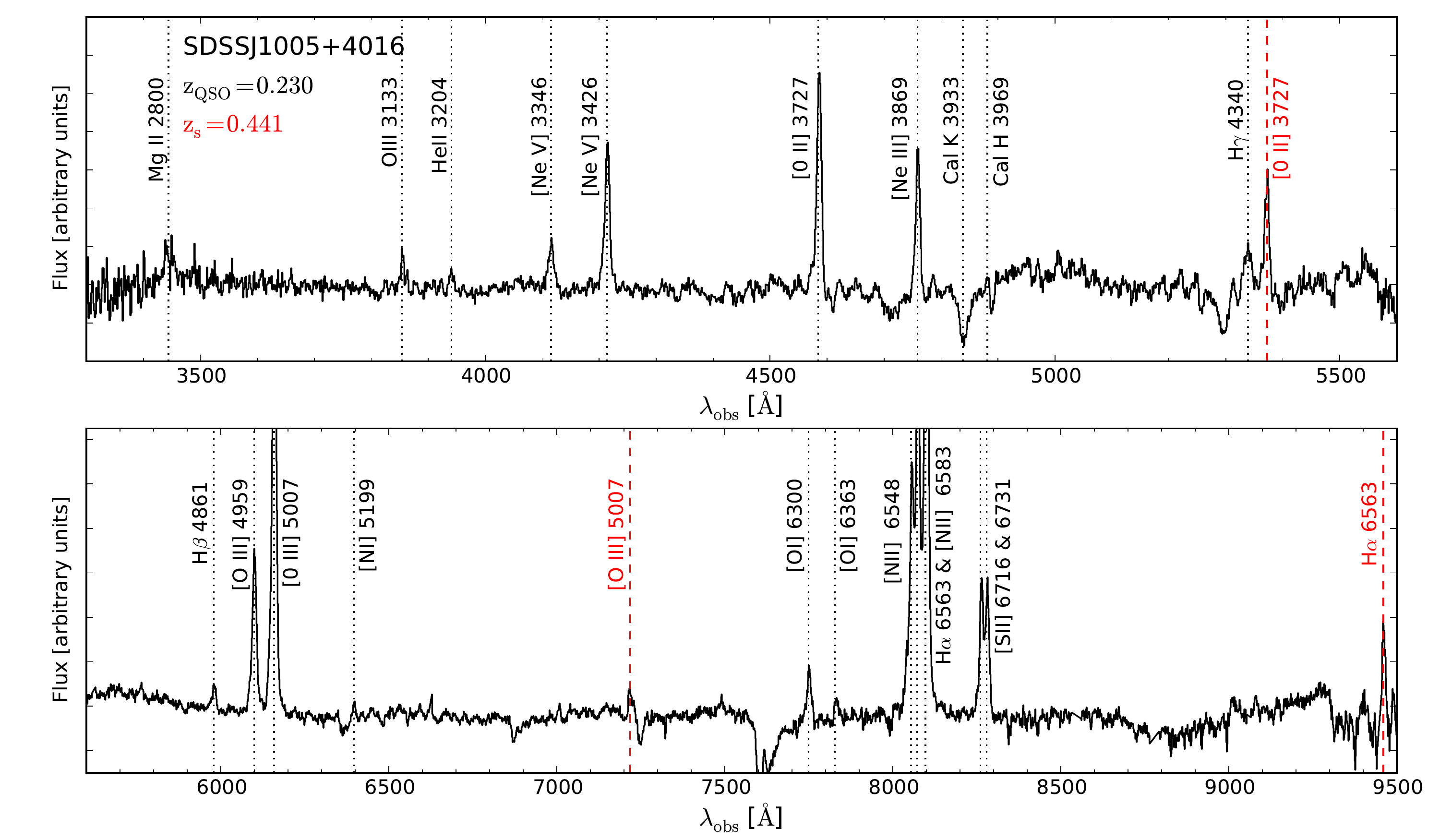}
\caption{Keck/LRIS spectra of the three new QSO lenses. From top to bottom are 
 \obja, \objb, and \objc. The emission lines are labelled as in Fig.~\ref{fig:SDSS_spectra}. 
For each object the two panels correspond to the spectra obtained in the blue and red channels of the 
LRIS spectrograph. All spectra are shown in the observed frame. They have not been corrected for
telluric absorption.}
\label{fig:LRIS_spectra}
\end{center}
\end{figure*}

\section{Object selection in SDSS}

Our search technique is similar to the one used to build the SLACS sample \citep{Bolton2008} and was 
originally proposed by \citet{Warren96}. It consists of selecting lenses following their spectroscopic properties
and looking for emission lines redshifted beyond the redshift of the potential lens. 
SLACS selects early-type massive galaxies as lenses. We select QSOs. 

We use the SDSS DR7 catalogue \citep{Aba2009} of spectroscopically confirmed QSOs 
 to build a sample of potential lenses, without applying any magnitude cut. We consider all 22,298 QSOs  with  $z_{\rm QSO} < 0.7$ 
  and with SDSS spectra.
   In these spectra, we look for emission lines redshifted beyond the redshift of the QSO. Since the SDSS fiber is comparable in size  (3\arcsec\, in diameter) to the typical Einstein 
radius of our QSOs, the object responsible for the background emission lines is very likely to 
be strongly lensed \citep[e.g.,][]{Bolton2008}.

The search technique itself consists in cross-correlating the QSO spectra with templates of 
emission line galaxies. The cross-correlation is carried out after removal of the QSO continuum. 
We choose very simple templates, with no continuum but which contain
the optical \OII\ doublet, the H$\beta$ hydrogen line, the \OIII\ doublet, and the H$\alpha$ line.
The best candidates selected by this technique, applied to SDSS spectra, show four clear emission lines
with a redshift larger than that of the selected QSO. One of these objects is 
\obj\ ($z_{\rm QSO}$=0.120; $z_{\rm s}$=0.641) as described in \citet{Courbin2010b}, where $z_{\rm s}$ is the 
redshift of the lensed background source. The three new objects studied in this paper are 
\obja\ ($z_{\rm QSO}$=0.293; $z_{\rm s}$=0.412), \objb\ ($z_{\rm QSO}$=0.209; $z_{\rm s}$=0.558), 
and \objc\ ($z_{\rm QSO}$=0.230; $z_{\rm s}$=0.441).  Fig.~\ref{fig:SDSS_spectra} shows these three SDSS spectra, 
where the emission lines of the background object are marked in red.

Note that the present work is a pilot study aimed at discovering some of the objects with fairly low lens-source contrast and with as many background emission lines as possible. In doing this, we highly bias the "sample", both in redshift and in lens mass. In particular, the range of source redshifts is restricted to $0.4<z_{\rm s}<0.7$ in order to see four lines simultaneously. The selection function within this range is complicated due to the sky emission lines in the red part of the spectrum and to the iron lines in the foreground QSO.  A detailed analysis of the selection window of our technique is well beyond the scope of this paper and depends on the adopted trade-off between the desired number of emission lines detected in the background sources and their detection level, i.e., the contrast between the source and the QSO. In addition, this contrast depends strongly on the relative redshifts of the QSO lens and the source.

\section{Keck spectroscopy and imaging}

We obtained spectroscopy of all three QSO lenses on
UT 2010 March 12 with the dual-beam Low Resolution Imaging Spectrometer,
LRIS \citep{Oke95, Rockosi2010}, on the Keck I telescope.  Two dithered
exposures  of each source were obtained.  The night was photometric, 
with seeing of $\sim 1\farcs2$.  The observations used the
1\farcs5 wide longslit, the 5600 \AA\, dichroic, the 400$\ell$~mm$^{-1}$
grating on the red arm of the spectrograph (blazed at 8500 \AA ;
spectral resolution $R \equiv \lambda / \Delta \lambda \approx 700$
for objects filling the slit), and the 600$\ell$~mm$^{-1}$  grism
on the blue arm of the spectrograph (blazed at 4000 \AA ; 
$R \approx 1000$).  This configuration allows simultaneous
coverage of essentially the entire optical window.

We observed SDSS~J0827+5224 for a total integration time of 600~s
at position angle $PA=+34.1^\circ$.  We observed SDSS~0919+2720 for a
total integration time of 900~s at $PA=+60.0^\circ$,
and we observed SDSS~1005+4016 for a total integration time of
1200~s at $PA=-10.0^\circ$.  Note
that LRIS has an atmospheric dispersion corrector, so there are no
issues with light lost due to observing at non-parallactic angles.
The data were processed using standard procedures, and flux calibrated
using archival observations of standard stars from \citet{Massey90}. 
The reduced spectra are presented in Fig.~\ref{fig:LRIS_spectra}.

The imaging observations were obtained on the night of UT 2010 May 17,  at the 10-m Keck I 
telescope in variable conditions. No flux calibration was obtained. 
Each object was observed in the $B$ and $R$ filters, with a pixel scale of 0.135\arcsec. The exposure 
times, seeing, observational setup and airmass for all Keck observations are summarized in Table~\ref{tab:summary}.

\begin{table}[t!]
  \caption{Summary of the Keck/LRIS  observations.}
  \vspace{0.2cm}
  \begin{tabular}{lrccc}
    \hline
    \hline
    Object & exp. time & Filter &        seeing       &  airmass  \\
                & (sec)         &                   &      ($\arcsec$) &                   \\
    \hline
    \obja   &       600     &       B            &      0.81       & 1.39    \\
                &       600     &       R               &  0.90       &  1.39      \\
                &       600     &   G400 red     &   1.20        & 1.19    \\
                &       600     &   G600 blue      &  1.20       &  1.19     \\
    \objb   &       600     &       B             &      0.75       &  1.13 \\
                &       600      &       R           &      0.75        &  1.13\\
                &       900      &   G400 red      &   1.20       &   1.01   \\
                &       900     &   G600 blue    &    1.20      &    1.01  \\
    \objc   &       600      &         B            &    0.75       &   1.12  \\
                &      600       &        R               & 0.70       &    1.12     \\    
                &     1200     &   G400 red           &  1.20     & 1.07        \\    
                &     1240     &   G600 blue         &  1.20     & 1.07        \\                                    
    \hline
  \end{tabular}
  \label{tab:summary}
\end{table}

\begin{figure}[t!]
\begin{center}
\includegraphics[width=9.cm]{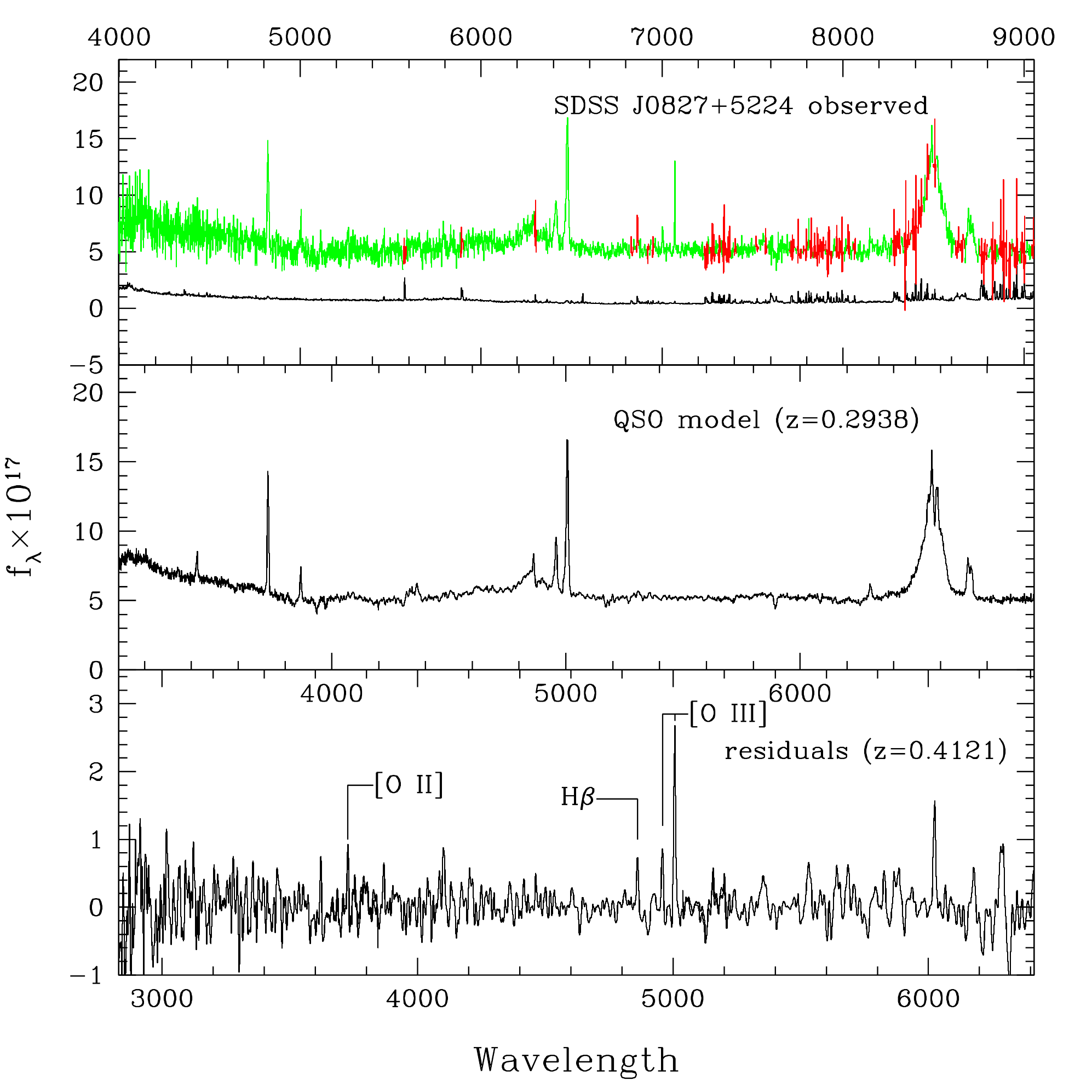}
\caption{Example of a spectral principal component analysis of a SDSS spectrum 
{\it Top:} SDSS spectrum of \obja\ along with the 1$\sigma$ errors (black). The spectrum is shown
in the observed frame, labeled on the upper axis.  The red points are those flagged as bad in the SDSS archive, 
typically because of poor sky subtraction.  {\it Middle:} Model of this spectrum, computed using the 43 most 
significant eigenspectra derived from 1000 high signal-to-noise spectra from the SDSS archive.  The model is 
plotted against rest wavelength of the QSO, labeled below the lower boundary of this panel. {\it Bottom:} Subtraction 
of the model spectrum from the observed one, smoothed with a Gaussian having a full width at 
half maximum $FWHM=200$~\kms.  The [OII], [OIII], and H$\beta$ emission lines from the lensed object are labeled.  
This difference spectrum is plotted against rest wavelength in the frame of the {\it lensed object}.  
The significant peaks that are not labeled are all associated with regions of poor sky subtraction.}
\label{fig:PCA_0827}
\end{center}
\end{figure}

\begin{figure}[t!]
\begin{center}
\includegraphics[width=9.5cm]{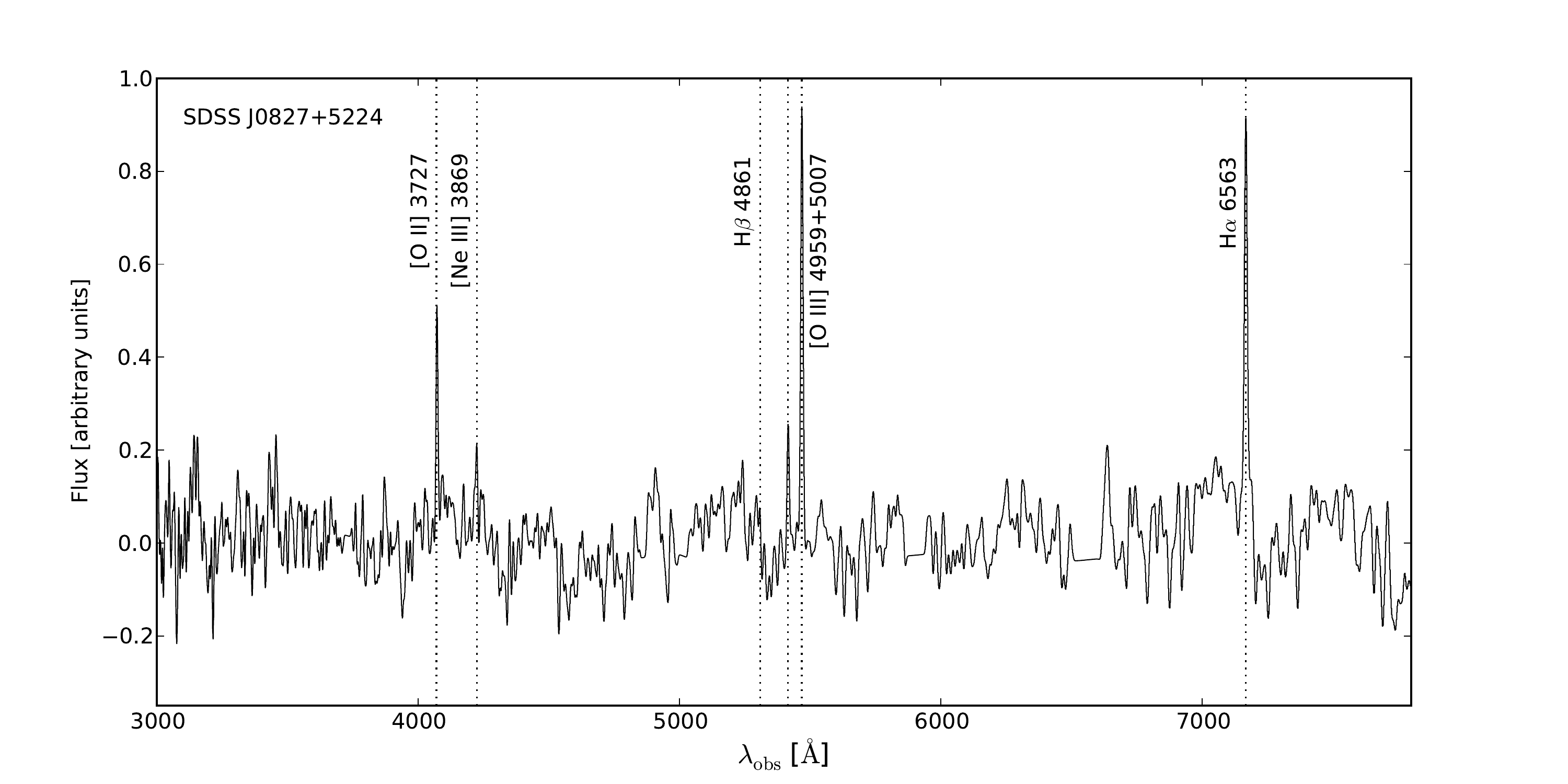}
\includegraphics[width=9.5cm]{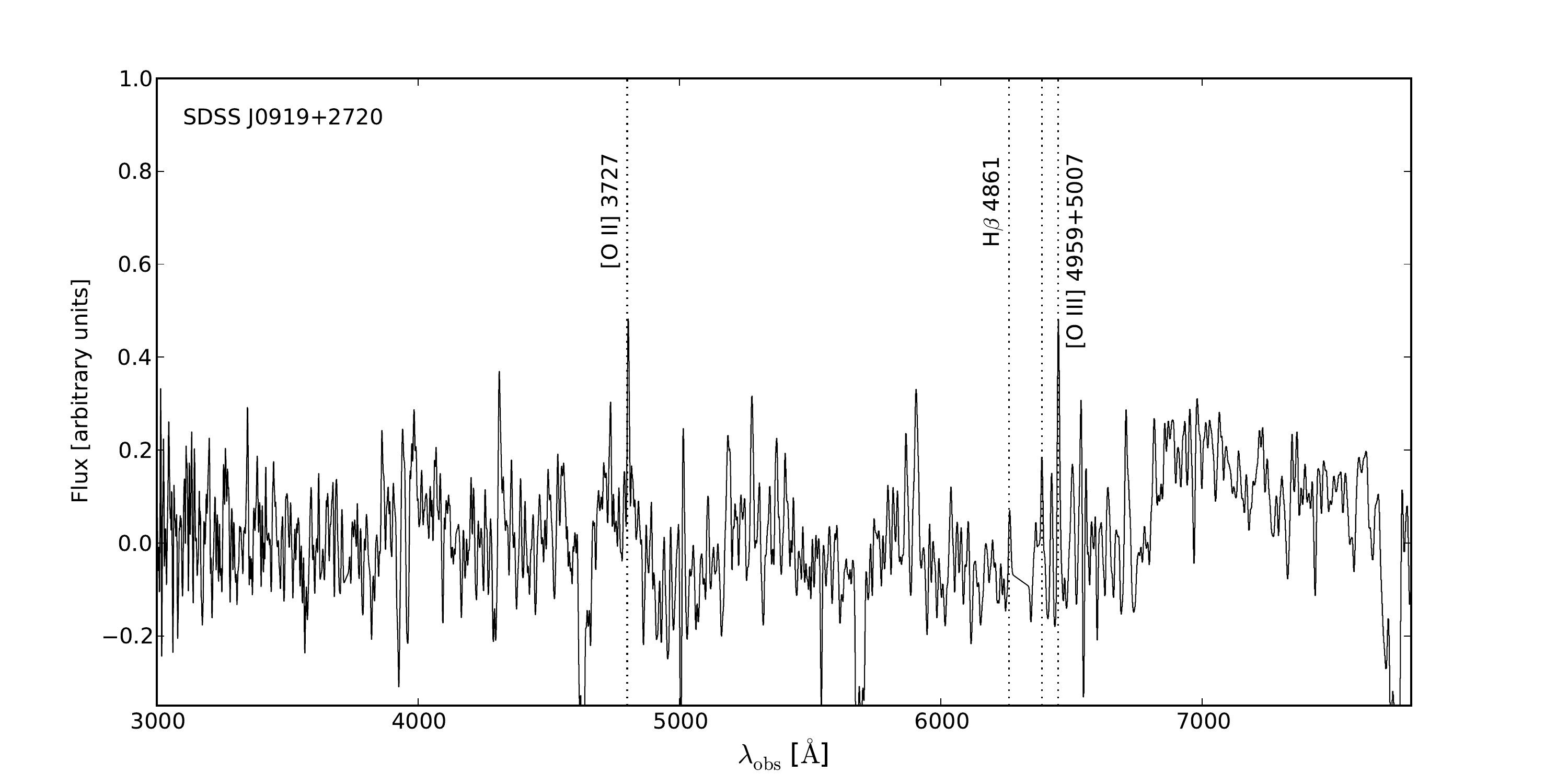}
\includegraphics[width=9.5cm]{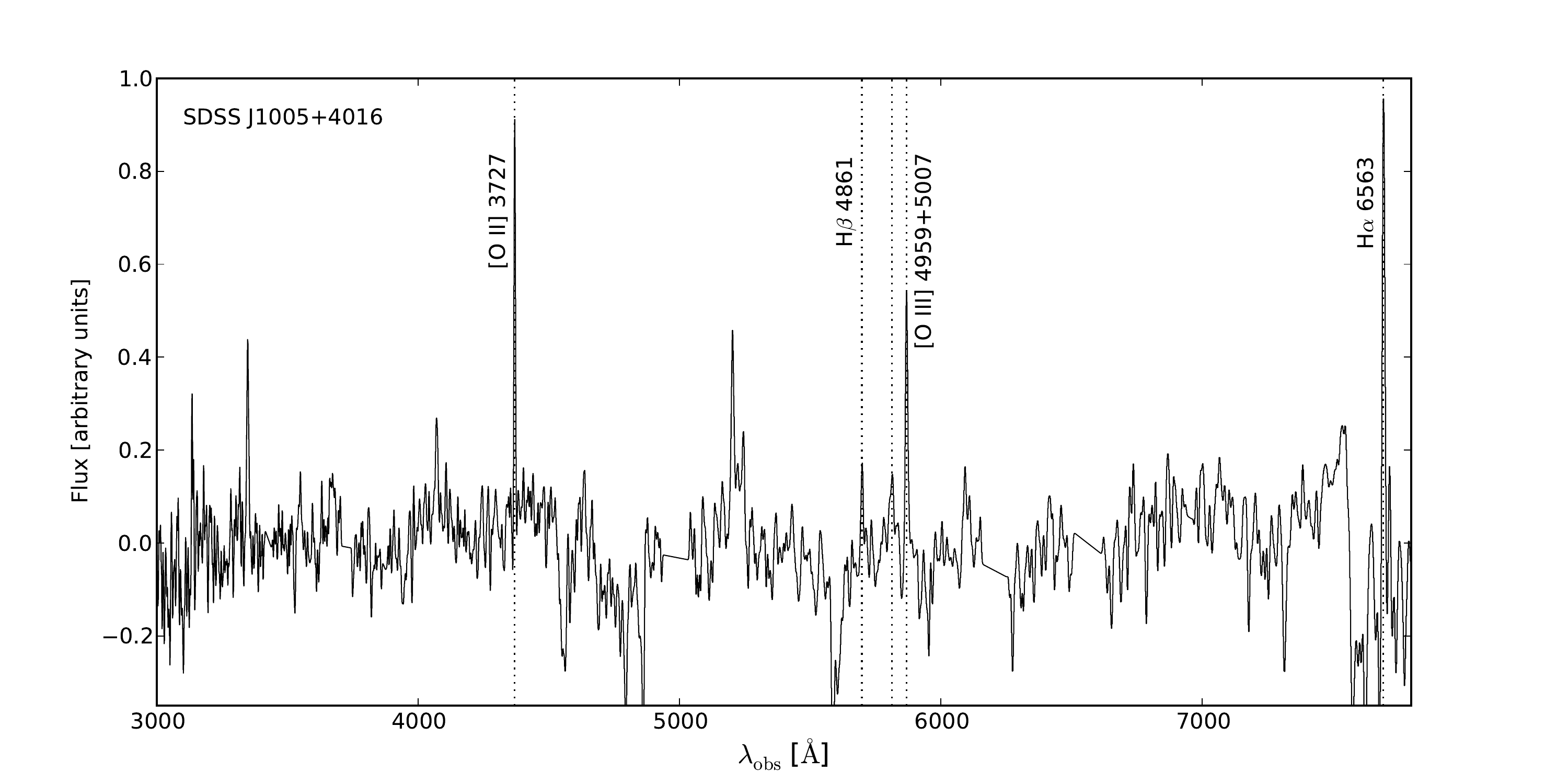}
\caption{PCA analysis of the Keck I spectra, shown in the observed 
frame. Between 3 and 6 of the emission lines of the background source are well detected
depending on the object.}
\label{fig:PCA_Keck}
\end{center}
\end{figure}

\section{Spectral principal component  analysis}

The spectral principal component analysis (PCA) procedure developed by \citet{Boroson2010} can be used 
to model any QSO spectrum.  This technique uses spectra of a large, representative sample of objects to 
generate eigenspectra, which can be fit to a given spectrum to get weights.  A linear combination of these 
eigenspectra can reproduce all the features of the input spectrum to within the noise of the observation,
given a sufficiently large set of eigenspectra constructed from spectra with high signal-to-noise.  
However, features that are not properly part of the object spectrum will not be represented in the 
eigenspectra, and so will not appear in the reconstruction.  Thus, the difference between the 
observed spectrum and the reconstructed model will show residual features due to a superposed 
object at a different redshift.  We used this technique as an alternative method to view the lines 
from the lensed objects relative to the noise and to search for weaker lines from the lensed objects.  
This worked well for the SDSS spectra where we had a large number of QSOs with which to create 
the eigenspectra.  An example is shown in Fig.~\ref{fig:PCA_0827}, where the spectrum of SDSS J0827+5224 is
modeled and subtracted, showing the residual lines from the lensed galaxy in the bottom panel.  
We did not detect any additional features from the lensed galaxies by this technique, but we plan 
to explore the use of this procedure to search for additional examples of objects lensed by QSOs 
within the SDSS archive. The main advantage of this approach over other techniques is that the many faint iron
features present in the spectrum of the foreground QSO are well removed, hence avoiding the
misidentification of emission lines from the background object with foreground iron emission. When applied to
our Keck spectra, the PCA analysis shows prominent emission lines from the lensed sources, well above
the noise level (Fig.~\ref{fig:PCA_Keck}). We  plan to use this technique to 
find objects with high source/lens contrast, allowing us to expand our sample of these
difficult to identify systems.

As we used this technique only to isolate features arising in the lensed object in previously discovered systems, we leave a complete analysis of its potential sensitivity to our future survey.  However, we note that \citet{Boroson2010} show that a spectral PCA model is able to reproduce the intrinsic spectrum with much greater fidelity than the input data typically do; thus, the significance of a feature detected in this way generally depends only on the signal-to-noise of the input spectrum, not on the quality of the model.

\section{Evidence for strong lensing from the Keck data}

The background emission lines detected in the SDSS spectra of our three QSOs are confirmed 
with the Keck spectra. With the longer wavelength coverage of the LRIS spectrograph, 
additional lines are even detected, like the strong H$\alpha$ emission of a galaxy in the background of
\objc. 

There is therefore little doubt that all three targets consist of a low redshift QSO almost 
perfectly aligned with a background emission line galaxy. Whether these background galaxies
are strongly lensed or not depends on the surface mass density of the foreground QSO host
galaxies. 

\begin{figure}[t!]
\begin{center}
\includegraphics[width=8.9cm]{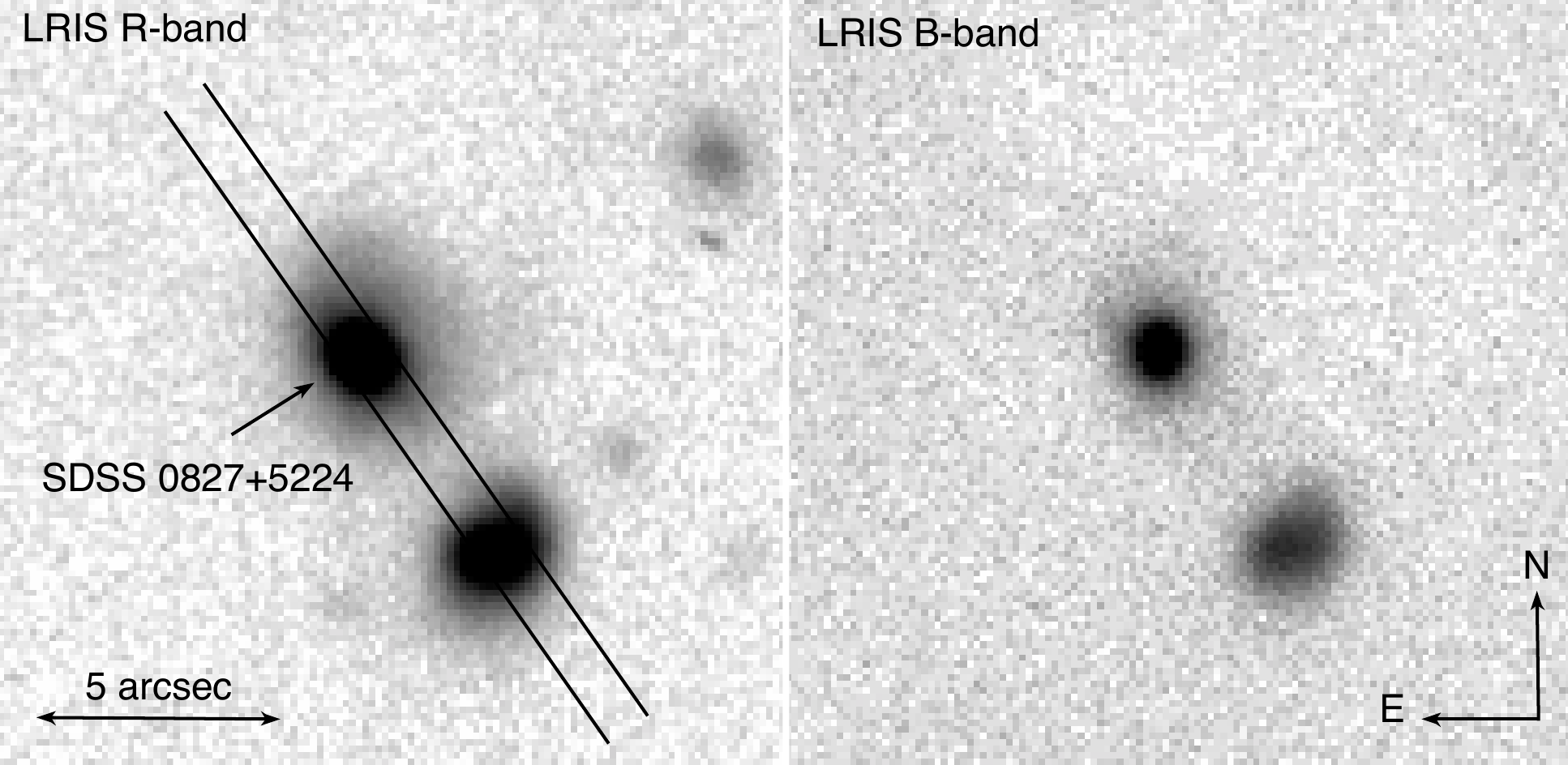}
\caption{Keck/LRIS images of \obja\, ($z=0.293$) in the $R$ and $B$ bands. 
The 1.5\arcsec\ slit of the LRIS spectrograph is shown, oriented with $PA=+30^\circ$. }
\label{fig:j0827_image}
\end{center}
\end{figure}

We use our Keck optical images to look for lensed multiple images or arcs within
3\arcsec\ of the QSOs. Due to the brightness of the QSOs this is a challenging task with
the spatial resolution of ground-based data. Image ``deconvolution" is required. We apply the 
MCS deconvolution software \citep{MCS98} to the data of \objb\ and \objc. No PSF star is 
available in the field of view of \obja, so no reliable deconvolution is possible for that object.
The images are shown in Figs.~\ref{fig:j0827_image}-\ref{fig:j1005_image}. When the deconvolution 
process is possible, the achieved resolution is  $FWHM=0.135$\arcsec. Although only the HST images 
presented in
Sect.~\ref{HST} allow to us to draw firm conclusions about all three objects, we find it interesting to 
briefly describe how the Keck and SDSS data alone already give convincing clues that these are strong lenses.

\subsection{\obja}

No image deconvolution is possible for this object. The images obtained in the two optical bands
(Fig.\ref{fig:j0827_image}) show no trace of any background object, in spite of the emission lines seen in the SDSS
spectrum. These lines become striking in our Keck
spectrum. This may be due to photometric variations of the source between the SDSS and the Keck observations 
or maybe due to aperture effects. Indeed, the 1.5\arcsec\ slit width of LRIS is twice as narrow as the 3\arcsec\ fiber of the SDSS 
spectrograph. The increased brightness of the background emission lines in the Keck spectrum, as compared 
with the SDSS spectrum, therefore suggests that the emitting object must be very close to the foreground QSO 
on the plane of the sky, too close to the QSO to be seen in our ground-based images.

From the Keck and SDSS data alone we conclude that we are in presence of two objects aligned with an 
accuracy of 0.75\arcsec\ (half the LRIS slit width) but  the background object is not spatially 
resolved in the Keck images. 

We note that the second object in the LRIS slit is a normal galaxy at the redshift of the QSO.

\subsection{\objb}

\begin{figure}[t!]
\begin{center}
\includegraphics[width=8.9cm]{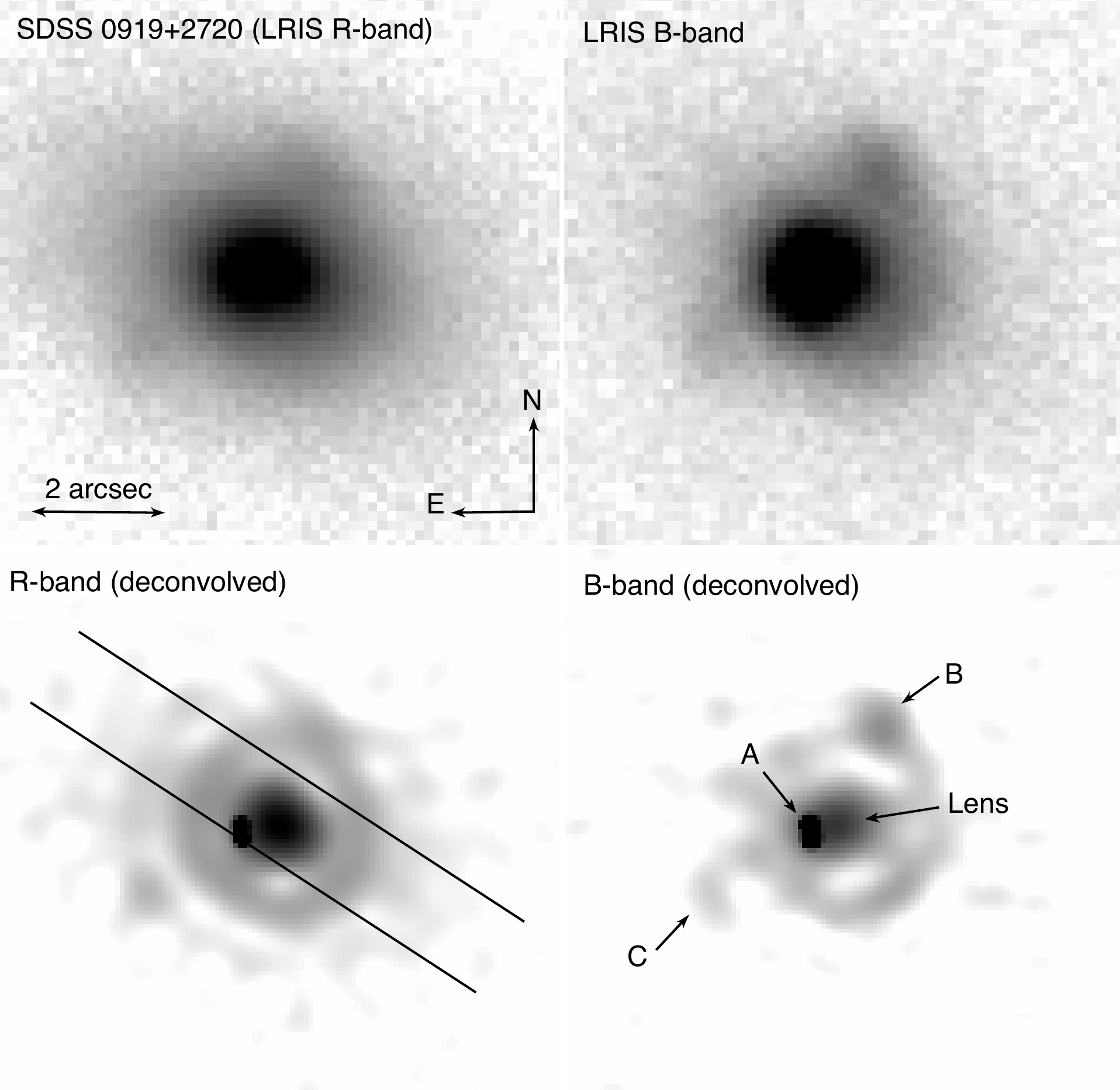}
\caption{{\it Top:} Keck/LRIS images of \objb\ ($z=0.209$) in the $R$ and $B$ bands. {\it Bottom:} Deconvolution of the 
two images unveiling a ring-like structure and a narrow pair of bright point sources, labelled A, that dominates
the total flux in the blue. In contrast, the object in the center of the ring, i.e., the putative lensing object,
is red. Two other objects are seen about the lens, labelled B and C. The 1.5\arcsec\ slit of the LRIS spectrograph is shown, and oriented with $PA=+60^\circ$. }
\label{fig:j0919_image}
\end{center}
\end{figure}

Several PSF stars are available near \objb. The deconvolved images (Fig.~\ref{fig:j0919_image}) 
reveal a complex system. Two objects are seen in the central arcsecond. One is extended and dominates
the total flux in the $R$ filter ("lens" in Fig.~\ref{fig:j0919_image}). The other is compatible with being a very 
narrow blend of two point sources that
dominates the flux in the $B$ filter ("A" in Fig.~\ref{fig:j0919_image}). The extended object is located at the 
center of a ring-like structure that 
makes a good candidate for a full Einstein ring. A plausible explanation is that the lensing galaxy is the red extended 
object, i.e., a low redshift early-type galaxy and that the point-like object is a QSO. The LRIS slit contains 
both objects. The integrated spectrum shows only two sets of lines at two different redshifts, not three. The QSO and the 
galaxy in Fig.~\ref{fig:j0919_image} must therefore be at the same redshift. 

Contrary to \obja, the background emission lines are seen stronger in the SDSS spectrum with respect to the
QSO lines than they do in the Keck spectrum. This is well explained by the slit orientation of LRIS, that catches
most of the light of the putative lens and QSO, but which misses most of the flux from the ring-like structure. The SDSS
fiber fully encompasses the ring, hence producing stronger lines. 

We conclude that we see an early-type galaxy lensing a background source as a full Einstein ring, and that
the lensing galaxy has a nearby QSO companion at the same redshift. It is not clear at this points whether the 
QSO has its own host galaxy or whether its host galaxy is the one at the center of the Einstein ring.
Although the QSO is brighter
than the galaxy, we expect it to be
much less massive and to contribute only as a perturber to the main lensing potential well. 
Structures are seen in the ring (B and C in Fig.~\ref{fig:j0919_image}).
It is not clear whether they are associated with the lensed source or whether they are unrelated objects seen in projection
along the line of sight. 

\begin{figure*}[t!]
\begin{center}
\includegraphics[width=17cm]{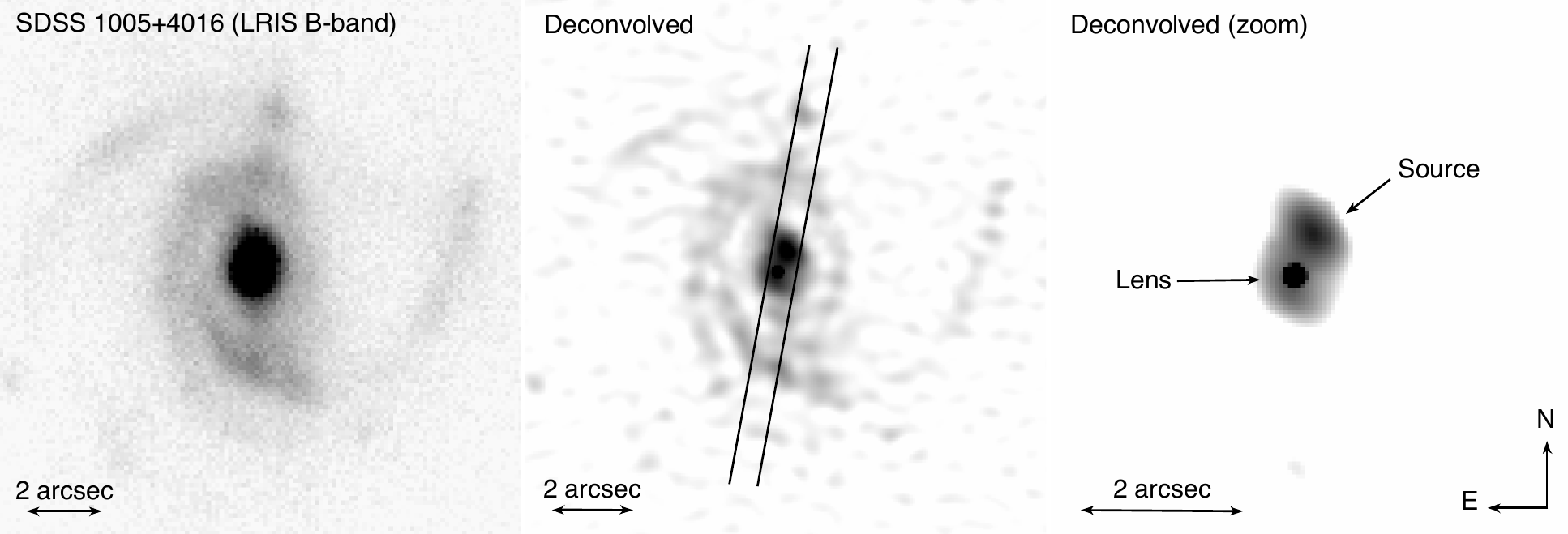}
\caption{{\it Left:} Keck/LRIS image of \objc\ ($z=0.230$) obtained in the $B$-band showing the face-on spiral  QSO
host galaxy. {\it Middle:} Deconvolved image showing that the nucleus of the spiral is 
double. The 1.5\arcsec\ slit of LRIS is shown in overlay, with $PA=-10^{\circ}$. {\it Right:} Zoom of the center of the 
deconvolved image showing the bulge of the foreground spiral and a second object $\sim$1\arcsec\ to the NW.}
\label{fig:j1005_image}
\end{center}
\end{figure*}

\begin{figure*}[t!]
\begin{center}
\includegraphics[width=18.2cm]{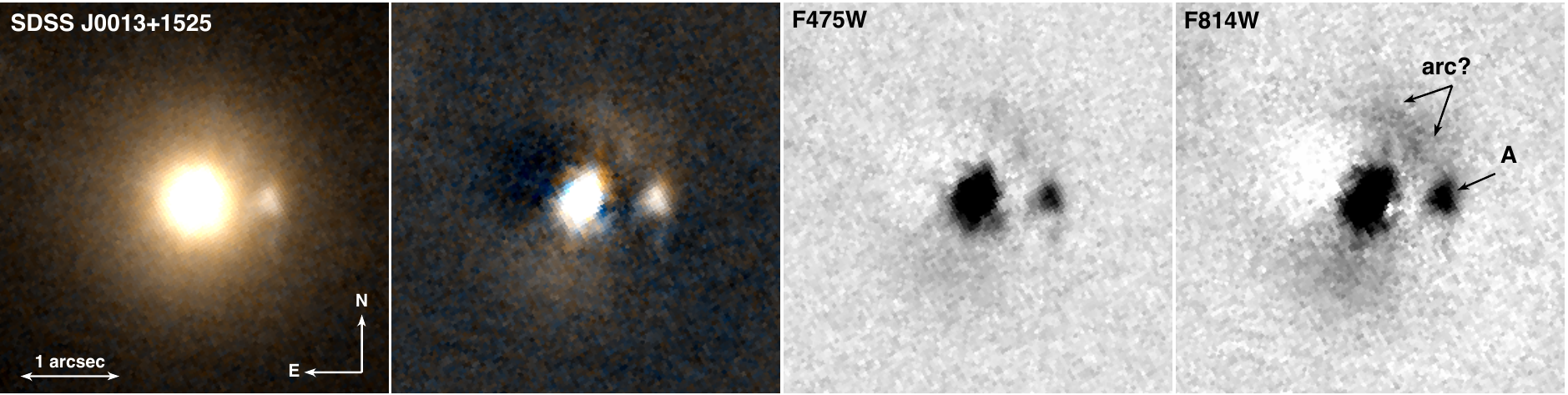}
\includegraphics[width=18.2cm]{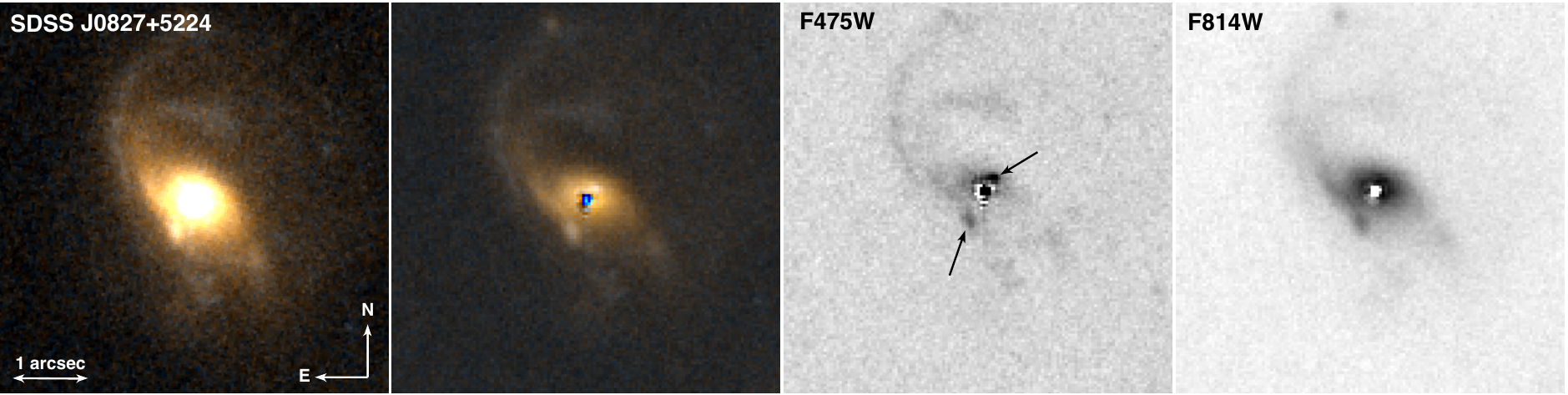}
\includegraphics[width=18.2cm]{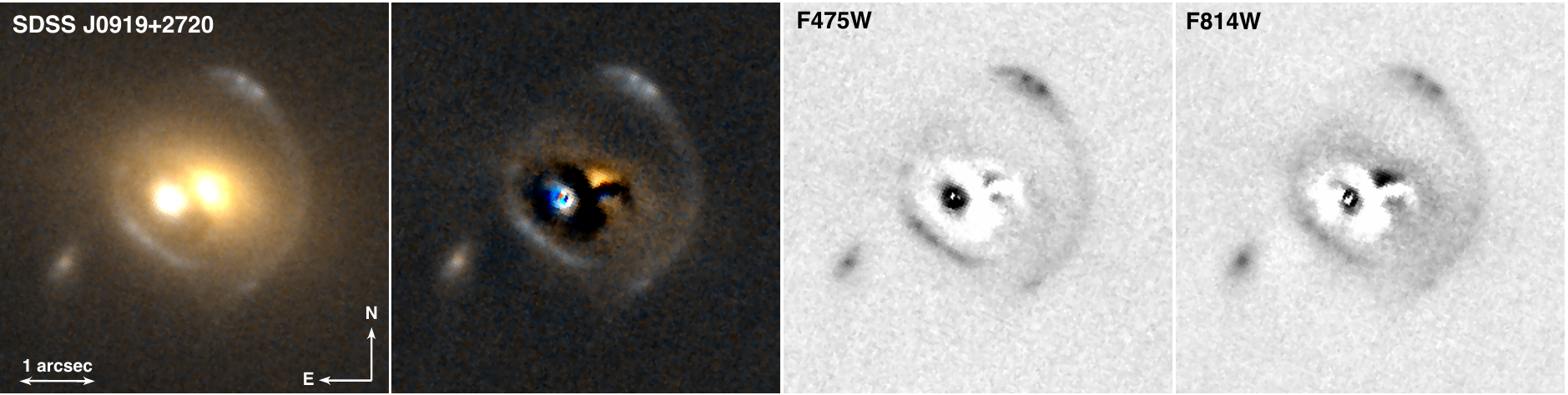}
\includegraphics[width=18.2cm]{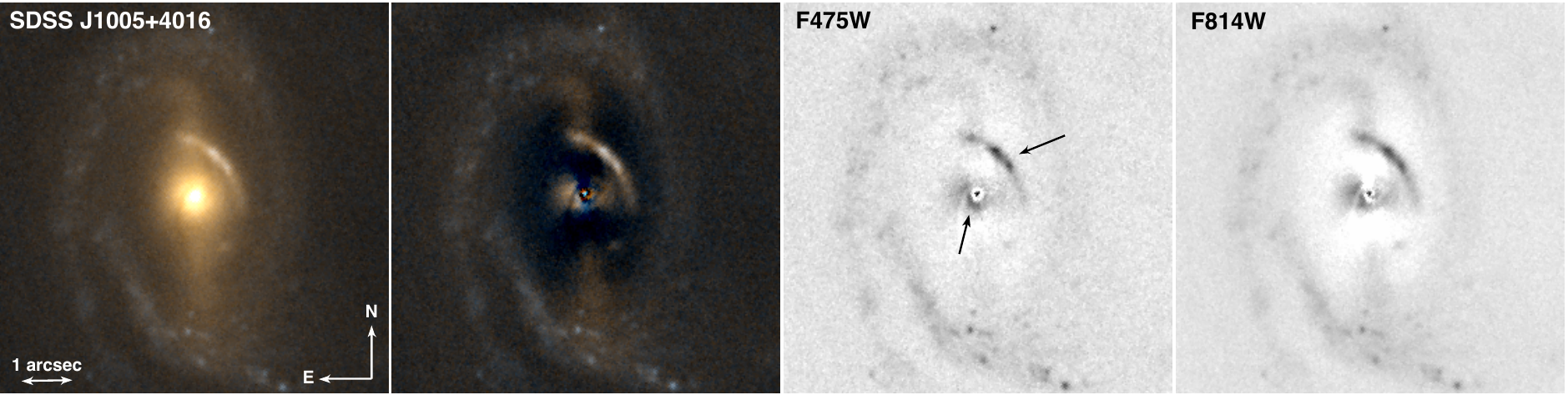}
\caption{HST/WFC3 images of our lensing QSOs. In each row, from left to right, we show (i) a color composite of 
the system, (ii) a  color composite after subtraction of the QSO and of the central parts of its host galaxy, (iii) the F475W
image after subtraction of the QSO and its host, (iv) the F814W image after subtraction of the QSO and its host. The arrows and 
labels indicate the lensed structures described in Section~\ref{results}. Note that \obj\ was already presented in \citet{Courbin2010a} but the present HST images do not clearly support the lensing nature of this object.}
\label{galfit}
\end{center}
\end{figure*}

\begin{table*}[t!]
\caption{Basic measurements for each object, including QSO and source redshifts, absolute magnitude of the QSO and of its host galaxy 
after subtraction of the QSO light. The measurements are given in the observed frame, in the Einstein radius. No k-correction is applied. (8) Einstein radius (9) total mass in the Einstein radius (10,11) 3-D stellar velocity dispersion and projected line of sight  velocity dispersion, as calculated from
our lensing model, assuming a circular, isotropic, and isothermal profile. }
\begin{center}
\def\arraystretch{1.2}
\begin{tabular}{c|c c c c c c c c c c  }
\hline
\hline
 Name & $z_{\rm QSO}$ & $z_{\rm s}$ & F475W & F814W  & F475W & F814W  & R$_{E}$ & M($< R_{E}$) & $\sigma_{v, 3D}$ & $\sigma_{v, LOS}$ \\
          &        	&        &   QSO    &    QSO   & Host      &  Host & (\arcsec) & (10$^{10} {\rm M}_\odot$)   & (km~s$^{-1}$) &  (km~s$^{-1}$)  \\
 \hline
 SDSS~J0013+1525 & 0.120 & 0.641 & -18.46 & -18.75&-18.75 & -20.15 & $<$ 0.74& $<$ 3.7 & $<$ 180.9  & $<$ 192  \\
 SDSS~J0827+5224 & 0.293 & 0.412 &  -19.95 &  -20.89 & -19.05 & -19.61 & 0.31 & $3.8\pm0.6$& $201^{+5}_{-9}$ & $213^{+5}_{-9}$   \\
 SDSS~J0919+2720 & 0.209 & 0.558 & -19.04 &-19.46 &-19.77& -21.80 & 1.22 & $21.4\pm6.7$& $268^{+9}_{-11}$ & $284^{+9}_{-11}$     \\
 SDSS~J1005+4016& 0.230 & 0.441 & -17.29   & -18.86  & -19.06 &-21.15 & 0.61 &  $7.5\pm1.1$ & $217^{+7}_{-9}$& $230^{+7}_{-9}$  \\
\hline
\end{tabular}
\end{center}
\label{measurements}
\end{table*}%

\subsection{\objc}

The Keck images of \objc\ show a face-on spiral galaxy with a bright nucleus.  Image deconvolution reveals 
a double nucleus, as shown in Fig.~\ref{fig:j1005_image}. One of the two "nuclei" is compatible with being a point 
source (QSO) superposed with an extended component, i.e., the nucleus of the spiral galaxy. The other, located
about 1\arcsec\ away from the QSO is elongated and not at the center of the spiral. If it is lensed
by the QSO+bulge of the galaxy, it does not show any counter image symmetric about the QSO. This object
is either not lensed or forms a partial Einstein ring. The HST images presented below demonstrate that the latter
hypothesis is correct.

\section{Confirmation with HST/WFC3 imaging}
\label{HST}

Our Keck images and spectra show strong evidence for lensing.
However a detailed analysis of the lensed images and  of the QSO host galaxy requires high resolution space images.
We have obtained such HST data (program GO\#12233), with the Wide Field  Camera 3 (WFC3) and the UVIS detector.


\subsection{Observations}

For each of the three objects, as well as for \obj, previously published in \citet{Courbin2010a}, we 
took six dithered exposures in the F814W filter and six exposures in the F475W filter. 
These observations correspond to one full orbit per filter and per object.

In addition to the QSO observations, we observed the open cluster NGC~136 in order to build a high 
signal-to-noise model of the PSF at the position of the QSO. This cluster contains stars with colors
similar to the QSOs and well sample the field of view, while still avoiding heavy crowding. The brightness of the stars
in this cluster also allow to apply exposure times that avoid strong saturation, even with the  overhead times
of  WFC3 due to transfer of  data frames in the camera buffer. NGC~136 was observed 
for one full orbit, using three long exposures (350 s) and one short exposure (60 s) for each filter. 

\subsection{Subtraction of the QSO and its host galaxy}

We carry out a two dimensional decomposition of the data using the {\tt Galfit} software \citep{peng2002, peng2010}, which is 
sufficient to identify which structures belong to the QSO+host and which can be
associated with the lensed source. For this purpose we use the pipeline-drizzled images provided by STScI.

For each object, we first build a PSF in the two filters using the observations of NGC~136. This is 
done by building a color-magnitude diagram for the cluster and  selecting stars with colors similar to 
our QSOs. The stars are also required to span a range of luminosities allowing to model both the core and the 
wings of the PSF.

We then fit a two-dimensional QSO+galaxy model convolved with the PSF and subtract it. The specific choice of 
the model is not important for the current application, as we only want to subtract as much light as possible from the
foreground object to unveil the background source. However, the best fits are obtained in all cases by using a 
combination of one PSF plus one or two Sersic profiles. The results of this analysis is displayed in Fig.~\ref{galfit}.

\section{Results}
\label{results}

The HST images immediately confirm the lensing nature of \objb, which displays a prominent 
blue Einstein ring surrounding
two bright objects. One of them is a point-source, i.e., the foreground QSO,
and the second, at the very center of the Einstein ring, is an early-type galaxy. The data reveals a dark extended 
structure centered on the QSO as well as a dark lane going through the galaxy. It is therefore likely that the lensing
object is in fact a QSO+galaxy merger at $z=0.210$ with significant amount of dust.

The second striking object is \objc. The elongated object seen in the deconvolved Keck images is beautifully confirmed 
as a partial lensed ring.
A close look at the residual image after subtraction of the lensing QSO reveals a small object about 0.1\arcsec\ to the
South-East of the QSO. This object is either produced by dusty structures in the nucleus of the spiral or a counter image 
of the arc. Detailed lensing models will tell whether it is a counter image of the arc of if we are observing a "naked cusp"
arc. 

\obja\ is also confirmed as a lens from the HST images. Two slightly extended sources are seen on each side of the central
QSO after proper PSF subtraction (shown with arrows in Fig.~\ref{galfit}). These objects are seen both in the F814W and in the F475W images, but the contrast with the QSO 
makes them most 
visible in the bluer F475W filter. As suspected from the relative brightness of the background emission lines seen in the 
SDSS and Keck spectra, the two lensed images of the source are  very close to the foreground QSO.

Finally, \obj\ is a more difficult system to confirm as a gravitational lens. \citet{Courbin2010a} detected two  point-like sources 
symmetrical about the foreground QSO, from near-IR adaptive optics images. The fainter of these sources is not visible in the 
 optical HST images, either because it is heavily obscured by dust or because it was an artifact in the Keck adaptive optics images.
The brighter of the two sources, labelled A in Fig.~\ref{galfit},
 is confirmed. It shows an extended 
arc-like structure to the North that could consist of a partial Einstein ring though this structure is faint and affected by the foreground 
QSO host  galaxy. Therefore, with the present data set, we cannot conclude with certainty that \obj\ is a genuine lens.


Basic measurements are given for all four systems in Table~\ref{measurements}. The mass estimates are derived from a Singular Isothermal Sphere (SIS) lens model. In Table~\ref{measurements}  the error bar on the total mass, M($< R_{E}$), is of the order of 15\%. This 
is due to the arbitrary choice of an isothermal mass profile to determine the lens total mass, and to the fact that we approximate $R_{E}$ by assuming it is equal to the size of the Einstein rings seen in the HST images. The error is then propagated in the calculation of the line of sight velocity dispersion. The latter assumes spherical symmetry for the lens.
The model assumes that the counter-image 
of the arc in \objc\ is real, i.e., that the arc is not a naked-cusp. For \obj, using a single image of the source to constrain the model 
leads to an upper limit for the mass of the QSO lens. 

\section{Conclusion}

Our combined Keck spectroscopic and imaging dataset, as well as the HST/WFC3 optical images, allow us to confirm 
three new QSOs acting as strong
lenses of background galaxies. Two of them, \obja\ and \objc, have  face-on host galaxies and background lensed images that
probe the very central part of the lensing potential. Detailed lens modeling combined with dynamical information on the 
host galaxies (rotation curve and central velocity dispersion) may therefore be very effective in breaking the bulge-halo
degeneracy. The third object, \objb, is peculiar and consists of QSO+galaxy merger. Gravitational lensing will offer 
an opportunity to model accurately the mass distribution in the pair and possibly to discriminate between the 
mass of the galaxy and that of the companion QSO. The lensing models needed to study these three unique active galaxies is 
beyond the scope of the present discovery paper and will be detailed in future work. 

The three objects presented here are the ones with the strongest background emission lines in SDSS DR7. Fainter 
and more numerous lines may be detected with cleaner QSO subtraction using the Spectral Principle Component Analysis technique. 
There is therefore hope to build a statistically significant sample of "QSO lenses" to study how the properties of QSOs and 
their host galaxies change with host mass and possibly with redshift.

\begin{acknowledgements}
This study is supported by the Swiss National Science Foundation (SNSF). SGD and AAM acknowledge a partial support from the 
NASA grant HST-GO-12233.01-A, the NSF grant AST-0909182, and the Ajax Foundation. The work of D. Stern was carried 
out at Jet Propulsion Laboratory, California Institute of Technology, under a contract with NASA. D. Sluse acknowledges partial support from the German Virtual Observatory and from the Deutsche Forschungsgemeinschaft, reference SL172/1-1.

This work makes use of the 
data collected by the SDSS collaboration and released in DR7. Funding for the SDSS and SDSS-II has been provided by the Alfred P. Sloan Foundation, the Participating Institutions, the National Science Foundation, the U.S. Department of Energy, the National Aeronautics and Space Administration, the Japanese Monbukagakusho, the Max Planck Society, and the Higher Education Funding Council for England. The SDSS Web Site is http://www.sdss.org/.

The SDSS is managed by the Astrophysical Research Consortium for the Participating Institutions. The Participating Institutions are the American Museum of Natural History, Astrophysical Institute Potsdam, University of Basel, University of Cambridge, Case Western Reserve University, University of Chicago, Drexel University, Fermilab, the Institute for Advanced Study, the Japan Participation Group, Johns Hopkins University, the Joint Institute for Nuclear Astrophysics, the Kavli Institute for Particle Astrophysics and Cosmology, the Korean Scientist Group, the Chinese Academy of Sciences (LAMOST), Los Alamos National Laboratory, the Max-Planck-Institute for Astronomy (MPIA), the Max-Planck-Institute for Astrophysics (MPA), New Mexico State University, Ohio State University, University of Pittsburgh, University of Portsmouth, Princeton University, the United States Naval Observatory, and the University of Washington.

\end{acknowledgements}

\bibliographystyle{aa}
\bibliography{Articles}
\end{document}